\def \cm{~\rm{cm}}
\def \s{~\rm{s}}
\def \km{~\rm{km}}

\def \K{~\rm{K}}

\def \erg{~\rm{erg}}

\def \yr{~\rm{yr}}

\def \kev{~\rm{keV}}
\documentclass[12pt,preprint]{aastex}
\usepackage{natbib}
\usepackage{epsfig}
\usepackage{amsmath}                
\usepackage{amsfonts}               
\usepackage{amssymb}                

\begin{document}   

\title{X-Ray Emission from Planetary Nebulae Calculated by 1D Spherical Numerical Simulations}

\author{Muhammad Akashi\altaffilmark{1}, Noam Soker\altaffilmark{1},
Ehud Behar\altaffilmark{1}, and John Blondin\altaffilmark{2}}

\altaffiltext{1}{Department of Physics, Technion$-$Israel
Institute of Technology, Haifa 32000, Israel;
akashi@physics.technion.ac.il; soker@physics.technion.ac.il;
behar@physics.technion.ac.il}

\altaffiltext{2}{Department of Physics, North Carolina State
University, Raleigh, NC 27695 Email: john\_blondin@ncsu.edu}

\begin{abstract}
We calculate the X-ray emission from both constant and time
evolving shocked fast winds blown by the central stars of
planetary nebulae (PNs) and compare with observations. Using
spherically symmetric numerical simulations with radiative
cooling, we calculate the flow structure, and the X-ray
temperature and luminosity of the hot bubble formed by the shocked
fast wind. We find that a constant fast wind gives results that
are very close to those obtained from the self-similar solution.
We show that in order for a fast shocked wind to explain the
observed X-ray properties of PNs, rapid evolution of the wind is
essential. More specifically, the mass loss rate of the fast wind
should be high early on when the speed is $ \sim 300-700 \km
\s^{-1}$, and then it needs to drop drastically by the time the PN
age reaches $\sim 1000 \yr$. This implies that the central star
has a very short pre-PN (post-AGB) phase.
\end{abstract}

{\it Subject headings:} Subject headings: stars: mass loss $—$
stars: winds, outflows $—$ planetary nebulae: X-ray $—$ X-rays:
ISM

\section{INTRODUCTION}

In the last seven years people have been trying to understand the
nature and the origin of the extended, spatially resolved X-ray
emission detected in planetary nebulae (PNs) by the {\it Chandra}
X-ray Observatory (CXO), e.g., BD~+30$^\circ$3639 (Kastner et al.\
2000; Arnaud et al.\ 1996 detected X-rays in this PN with ASCA),
NGC 7027, (Kastner, et al.\ 2001), NGC 6543 (Chu et al.\ 2001),
Henize 3-1475 (Sahai et al. 2003), Menzel 3 (Kastner et al.\
2003),  and by the {\it XMM-Newton} X-ray Telescope, e.g., NGC
7009 (Guerrero et al.\ 2002), NGC 2392 (Guerrero et al.\ 2005),
and MGC 7026 (Gruendl et al. 2004). The X-ray emitting gas can
come from shocked fast wind segments that were expelled by the
central star during the late post-asymptotic giant branch (AGB)
phase and/or early PN phase, and/or the X-ray emitting gas may
result from a collimated fast wind (CFW), or jets, blown in
conjunction with the companion to the central star during the late
AGB phase or early post-AGB phase (Soker \& Kastner 2003). These
interactions play a significant role in shaping PNs (e.g., Balick
\& Frank 2002, and references therein). Therefore, understanding
the X-ray emission can teach us about the properties of the
central fast wind and CFW, and by that shed light on the shaping
mechanism of PNs.

One of the major properties of the X-ray emitting gas to be
understood is its relatively low temperature of $\sim 1-3 \times
10^6 \K$. The observed velocity of fast winds in PNs is $v_f>1000
\km \s^{-1}$, implying a post shock temperature of $T> 10^7 \K$,
much higher than observed in the X-ray emitting gas. Two solutions
are possible: (1)
The hot ($T> 10^7 \K$) post-shock gas is cooled via heat
conduction by interaction with the cold, slow-wind material, which
is at a temperature of $T \sim 10^{4} \K$ and most X-ray emission
comes from the conduction front (Soker 1994; Zhekov \& Perinotto
1996; Steffen et al. 2005). A similar effect can result from
mixing of hot post-shock fast-wind gas and cool slow-wind gas (Chu
et al. 1997). Instability modes near the contact discontinuity can
lead to significant mixing of cold and hot gas (Stute \& Sahai
2006, hereafter SS06). (2) The X-ray emitting gas comes mainly
from a slower moderate-velocity wind of $v_f \sim 500 \km
\s^{-1}$. This flow can be a post-AGB wind from the central star
(Soker \& Kastner 2003; Akashi et al. 2006 hereafter ASB06; and
SS06), or two opposite jets (or CFW, Soker \& Kastner 2003). The
idea of a post-AGB wind is supported by the analytical
calculations of a spherically- symmetric fast wind done by ASB06
and based on the self-similar solution of Chevalier \& Imamura
(1983). The self similar solutions, however, are limited in their
ability to account for temporal evolution and to properly treat
radiative cooling.

In the present paper, we carry out a numerical study for the
expected contribution from the fast spherical wind blown by the
central star. We run numerical simulations solving the
hydrodynamical equations with a full treatment of radiative
cooling, and by setting proper initial and boundary conditions of
the interacting winds, e.g., a fast wind evolving with time. Some
of the results obtained here are similar to results obtained very
recently by SS06. The remainder of the paper is outlined as
follows: The numerical method is described in \S 2. In \S 3 we
compare the spherical symmetrical numerical simulations with the
self-similar results of ASB06. In \S 4 we present several cases of
a time evolving fast wind. A short summary is given in \S 5.

\section{NUMERICAL METHOD}

The simulations were performed using \emph{Virginia}
\emph{Hydrodynamics-I} (VH-1), a high resolution multidimensional
astrophysical hydrodynamics code developed by John Blondin and
co-workers (Blondin et al. 1990 ; Stevens et al. 1992; Blondin
1994). The code uses finite-difference techniques to solve the
equations of an ideal inviscid compressible fluid flow. {{{ We use
1024 grid points to resolve the calculated region. The distance
between adjacent grid points increase as the flow expands. Using
more or less grid points (e.g., 512) changes the results only by a
few per cents or less.  }}} Radiative cooling was incorporated for
gas temperatures above $10^4 \K$, using the cooling function
$\Lambda (T)$ for solar abundances from Sutherland \& Dopita
(1993; their table 6). {{{ Radiative cooling should be treated
carefully near the contact discontinuity. The hot bubble cools
slowly because of its low density, while the dense shell cools
slowly because of its low temperature ($T=10^4 \K$) where the
cooling rate is low. The one or two grid points at the interface
between the hot bubble and cold shell could share high temperature
from the bubble and high density from the cold shell. Therefore,
radiative cooling at the interface may be overestimated. Actually,
low resolution in the code might in a sense mimic heat conduction.
Our code includes a check of this numerical problem. It takes for
the cooling rate at the contact discontinuity the minimum cooling
rate of the two zones (either from the cold shell or from the
bubble). }}}

To mimic the ionizing radiation of the central star, we did not
let the gas to cool to temperatures below $10^4 \K$ when radiative
cooling was included. For the initial conditions we take a
spherically symmetric slow wind with a constant mass loss rate
$\dot M_1$ and a constant velocity $v_1=10 \km \s^{-1}$. The slow
wind fills the space around the center from a minimum initial
radius $R_{c0}=R_c(t=0)$. This implies that the initial density is
$\rho_1=\dot M_1/4 \pi v_1 r^2$ for $r>R_{c0}$ and zero inside
this radius. At time $t=0$ a spherically symmetric fast wind with
a mass loss rate of $\dot M_2(t)$ and a velocity $v_2(t)$ is
turned on close to the center.

{{{ The reasons for setting up a vacuum in the center of the slow
wind before the fast wind is turned on are: (1) There is a large
uncertainty as to the evolution of the wind in the transition from
the AGB phase to the PN phase. Therefore, there is no clear
parameters to use for the wind at this stage. (2) This wind period
has relatively low mass, as mass loss rate is lower than in the
AGB phase. On the other hand, it is dense and slow enough to cool
very rapidly. So practically, it will form a shell inward to the
slow wind. As far as the numerical procedure is concerned, we can
mimic this intermediate wind segment by changing the value of
$R_{c0}$. As we show here, this does not change our results
much.}}}

\begin{figure}  
\resizebox{0.95\textwidth}{!}{\includegraphics{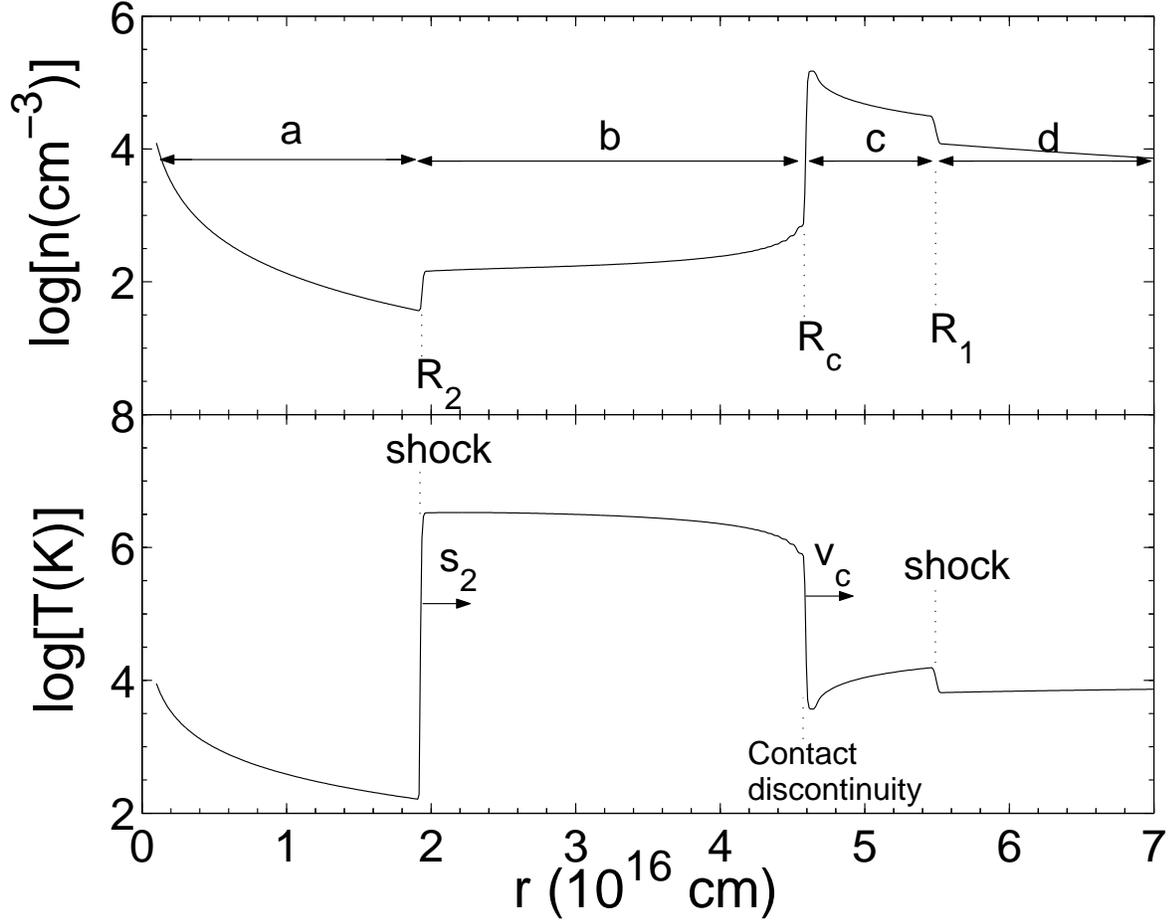}}
\caption{Total number density (upper panel) and temperature (lower
panel) at $t= 400 \yr$ from a spherical symmetric constant-flow
 simulation with no radiative cooling
included, and with the following parameters: Mass loss rate and
velocity of the slow wind: $\dot M_1(t)= 7 \times 10^{-6} M_\odot
\yr^{-1}$ and $v_1(t)=10 \km \s^{-1}$. Mass loss rate and velocity
of the fast wind: $\dot M_2(t)= 1.4 \times 10^{-7} M_\odot
\yr^{-1}$ and $v_2(t)=500 \km \s^{-1}$. The marked regions are:
($a$) undisturbed central star fast wind; (b) Hot bubble; (c)
Shocked slow wind; and (d) Unshocked nebular gas (AGB wind). $v_c$
and $s_2$ represent the velocities of the contact discontinuity
and inner shock, respectively. } \label{run1}
\end{figure}
The results of one run with a non-evolving fast wind of $\dot
M_2(t)= 1.4 \times 10^{-7} M_\odot \yr^{-1}$ and $v_2(t)=500 \km
\s^{-1}$, but without radiative cooling, is shown in Fig.
\ref{run1} at time $t= 400 \yr$. {{{ $R_{c0}=3\times 10^{15} \cm$
in this run. }}} In all runs $t= 0$ is when the fast wind starts.
In this run, $\dot M_1=7 \times 10^{-6} M_\odot \yr^{-1}$ (more
examples can be found in SS06.) The fast wind collides with the
slow wind and two spherical shock waves are formed. One shock
moves outwards at radius $R_1(t)$ into the undisturbed slow wind,
while an inner shock runs into the expanding undisturbed fast wind
at radius $R_2(t)$. Between the two shock fronts lies the contact
discontinuity at radius $R_c(t)$ (see also Fig.~2 by Volk~\& Kwok
1985). The values of the density and the temperature jump across
the contact discontinuity, while the velocity and pressure vary
continuously across it. The volume inside the dense shell $R_2 < r
< R_c$ (region $b$) in runs without radiative cooling is filled
with hot shocked fast wind. When radiative cooling is included,
the relatively dense region behind the contact discontinuity cools
radiatively and forms a cold layer at $T \sim 10^4 \K$ of
previously shocked fast wind material. The volume filled with hot
X-ray emitting gas is termed the hot bubble (region b). The
shocked slow wind in the shell $R_c < r < R_1$ (region $c$) cools
much faster due to its high density maintaining a temperature of
$\sim 10^4 \K$.

We are interested in the X-ray luminosity $L_x$ and temperature of
the hot gas. The luminosity is calculated in the energy band
$0.2-10 \kev$, as described in ASB06.  This range is chosen to
reflect approximately the sensitivity regime of the {\it Chandra}
and {\it XMM-Newton} telescopes and science instruments. As
discussed by SS06, for gas temperatures of $T_x \la 3 \times 10^6
\K$ most of the cooling takes place via emission at lower energies
($<0.2$~keV), which is outside the range of the instruments we
compare our results with. The temperature of the X-ray emitting
gas in the hot bubble varies with radius (see Fig.~\ref{run1}). In
ASB06 (eq. 9 therein), we introduced a mean emission-measure
weighted temperature $T_x$ that would be deduced from a
single-temperature model for the X-ray spectrum. Here too, we will
refer to the mean temperature of the hot bubble as $T_x$.

In the numerical model, the flow relaxes to a more or less steady
state after $\sim 100$~yr. For several hundred years, the velocity
of the contact discontinuity, $v_c$, and the velocity of the inner
shock, $s_2$, oscillate around their mean values.
However, the influence of these oscillations on the observed X-ray
luminosity and temperature is very small. At the first encounter
of the fast wind with the slow wind, the temperature of the X-ray
emitting gas is that of the post-shock gas. After $\sim 100$~yr,
as the post-shock gas expands with the nebula, the hot bubble
cools, and $T_x$ decreases as most of the radiation comes from the
region close to the contact discontinuity. Some other flow
properties at the very early stages of the winds interaction
process, are discussed by SS06. SS06 compare also the velocity of
the contact discontinuity in their numerical simulations with
analytical expressions. For the run presented in Fig. \ref{run1},
we find $v_c=36.2 \km \s^{-1}$. This run is referred to as model
B5 in ASB06 and in SS06, with the difference that ASB06 and SS06
took the undisturbed slow wind temperature to be very low, $T \ll
10^4 \K$, while here $T=10^4 \K$. The self-similar solution with a
cold slow wind ($T \ll 10^4 \K$) gives $v_c=32.6 \km \s^{-1}$. The
higher temperature here ($T = 10^4 \K$) implies more energy in the
flow, hence larger $v_c$. When SS06 take $T \ll 10^4 \K$ and
include radiative cooling, they find $v_c=27 \km \s^{-1}$, as a
consequence of the loss of energy by radiation. The actual
temperature of the undisturbed slow wind will rise from a few $100
\K$ during the pre-PN (PPN) phase to $\sim 10^4 \K$ during the PN
phase, after ionization starts. Neither we nor SS06 treat this
full evolution as we do not include photo-heating. We prefer to
take the temperature as appropriate for the PN phase, while SS06
take the PPN-phase temperature. This is not a major issue, as the
temperature of the pre-shock slow wind affects only slightly the
X-ray properties. On another matter, residual differences between
our results and those of SS06 can be attributed to slight
numerical differences in fitting the cooling function given by
Sutherland \& Dopita (1993).

\begin{figure}  
\resizebox{0.95\textwidth}{!}{\includegraphics{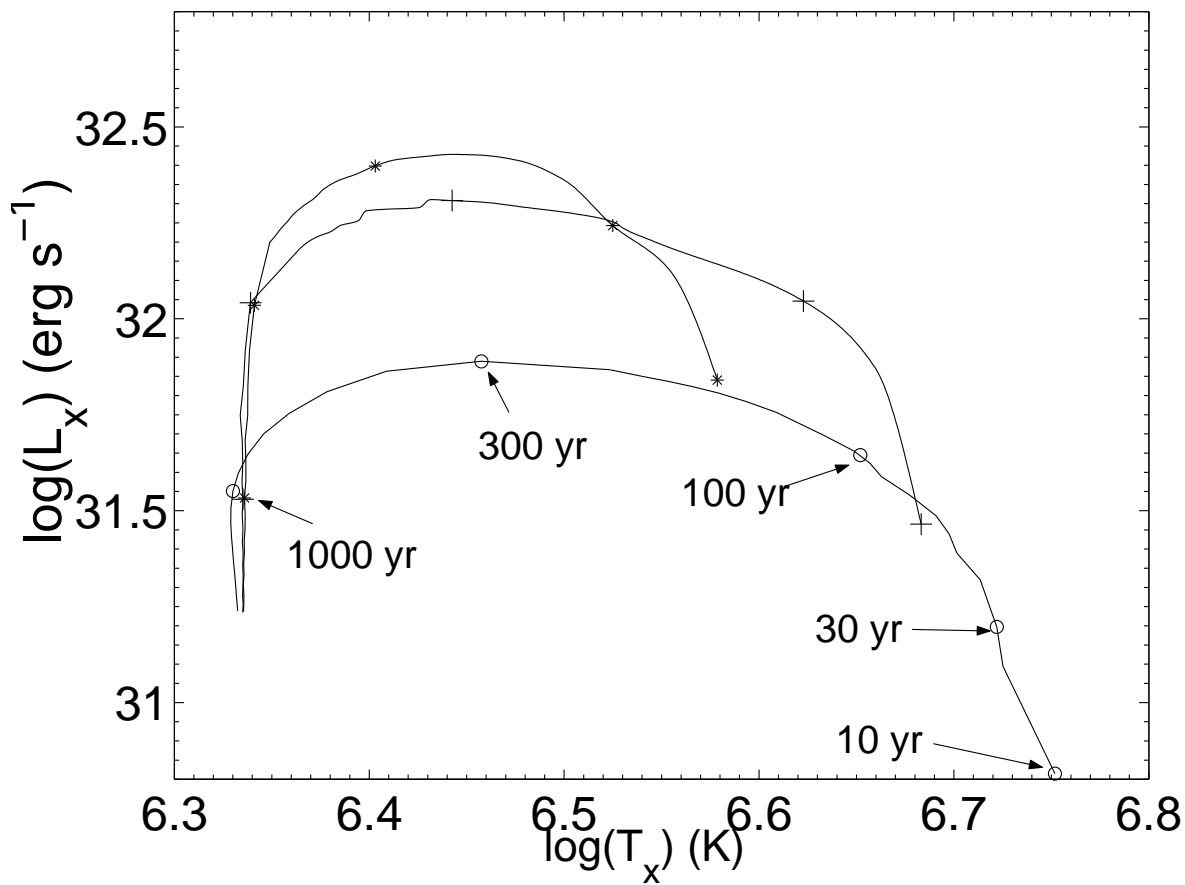}}
\caption{Simulated evolution of PN X-ray emission from $t=10$~yr
to $t=1000$~yr for three cases differing by their initial inner
radii of the slow wind: $R_{c0}=3 \times 10^{15} \cm$ (asterisks);
$R_{c0}= 5\times 10^{15} \cm$ (plus signs); $R_{c0}=1.2 \times
10^{16} \cm$ (circles). Data points for all runs are given at five
times as marked on the lower plot. In all runs $v_1(t)=10 \km
\s^{-1}$ (as in the rest of the paper), $\dot M_1(t)= 7 \times
10^{-6}  M_\odot \yr^{-1}$, $\dot M_2(t)= 1.4 \times 10^{-7}
M_\odot \yr^{-1}$, and $v_2(t)=500 \km \s^{-1}$. } \label{early}
\end{figure}

At early times, the X-ray emission of the PN depends strongly on
the slow wind initial radius $R_{c0}$. This is demonstrated in
Fig.~\ref{early}, which shows the PN evolution with time in the
$L_x-T_x$ plane for three different initial inner radii $R_{c0}$.
As explained above, the temperature is high at very early times,
and decreases toward its self similar solution at later times. At
early times the X-ray luminosity is very low as there is not much
gas in the hot bubble. With time, the X-ray luminosity increases
as more mass is accumulated in the hot bubble (SS06) until the
effect of expansion takes over, the temperature stabilizes and the
X-ray luminosity decreases as in the self similar solution (see
next section). To reach the self similar behavior takes more time
for larger values of $R_{c0}$, but regardless of $R_{c0}$ all
simulations converge to the same results as seen in Fig.
\ref{early}. Therefore, we will not discuss the very early times
of evolution. In most runs, those lasting 5000~yr, we took
$R_{c0}=8 \times 10^{15} \cm$. In the shorter runs, lasting
2000~yr, we took $R_{c0}=3 \times 10^{15} \cm$.

\begin{figure}  
\resizebox{0.95\textwidth}{!}{\includegraphics{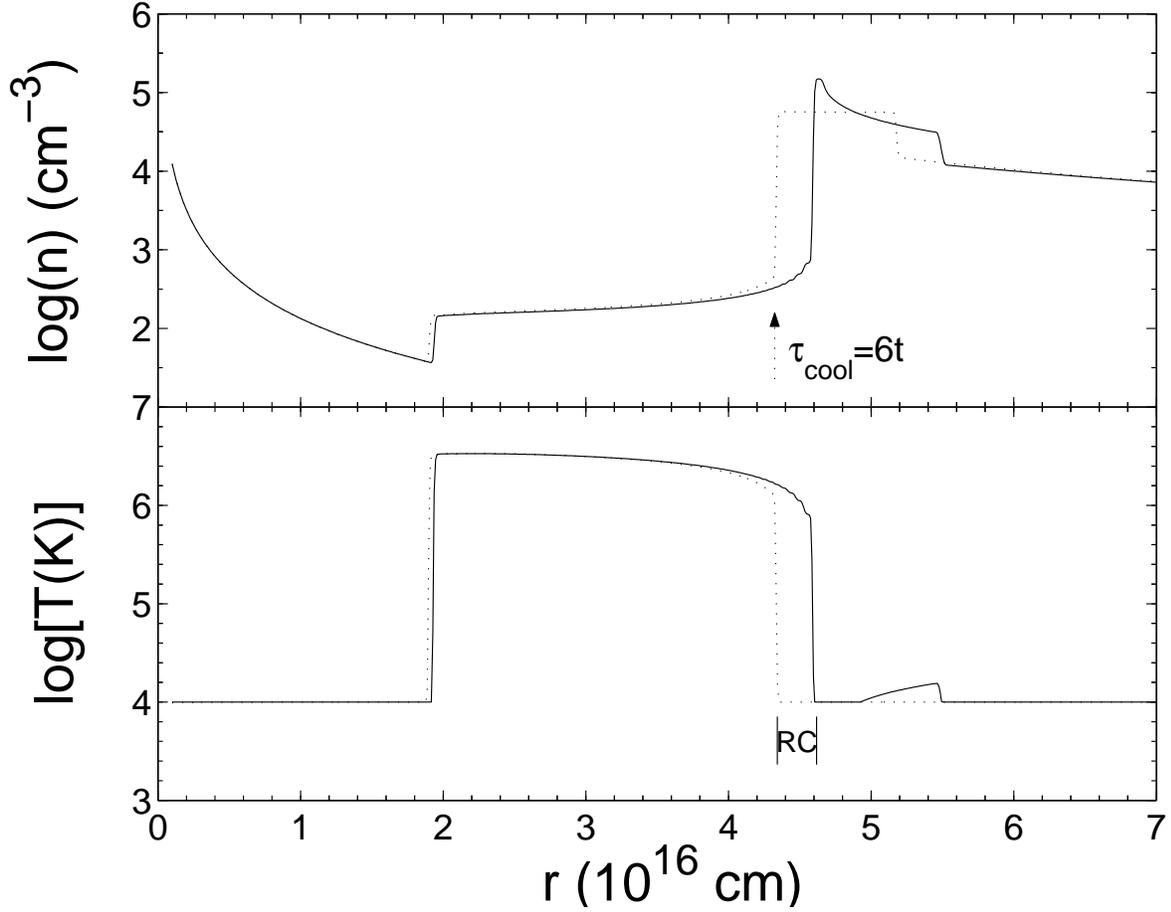}}
\caption{Simulation results for density and temperature profiles
with and without cooling. The solid lines are the same as in Fig.
\ref{run1}, except that $T= 10^4 \K$ was set as a strict lower
limit. The dashed lines show the results for the same parameters,
but with radiative cooling included. The region marked $RC$ is the
region where radiative cooling is most effective. It is found to
be the point where the radiative cooling time equals six times the
age of the flow. It is seen that radiative cooling reduces the
pressure in the hot bubble and hinders its expansion.}
\label{cool1}
\end{figure}

In order to demonstrate the effect of radiative cooling, we
present in Fig. \ref{cool1} a comparison between simulations with
and without radiative cooling. The effect of radiative cooling is
most pronounced at the densest and coolest shocked fast wind
segments, which reside right behind the contact discontinuity. As
explained above, no gas has a temperature below $10^4 \K$. In the
hot bubble, generally $10^5<T<10^7 \K$ and the radiative cooling
time $\tau _{\rm cool}$ is highly sensitive to the temperature
($\sim T^{5/2}$). This explains the sharp boundary of the
radiatively cooled region marked as $RC$ in Fig.~\ref{cool1}. {{{
We note that the region marked $RC$ is found to be the point where
the radiative cooling time equals six times the age of the flow.
Namely, regions with cooling time much longer that the flow time
had time to cool. The reason why these regions had time to cool is
that the gas now residing near the contact discontinuity (at
$R_c$) was shocked at much smaller radius, and hence its density
was higher and radiative cooling time much shorter in the past.
Therefore, we would expect that the present radiative cooling time
at $R_c$ is longer than the flow age. }}}

\section{CONSTANT FAST WIND}

We run several simulations with a constant, unevolving fast wind,
namely, $\dot M_2(t)$ and $v_2(t)$ do not vary with time. The
results enable a comparison with the self-similar results of ASB06
and serve as reference for the evolving fast wind simulations to
be discussed in the next section. Radiative cooling is included in
all runs in this section. We recall that in order to account for
radiative cooling in the self similar solutions, we simply
disregard wind segments with cooling times shorter than the PN age
($t_{\rm cool} <t$, see ASB06) in calculating the X-ray emission.

\begin{figure}  
\resizebox{0.95\textwidth}{!}{\includegraphics{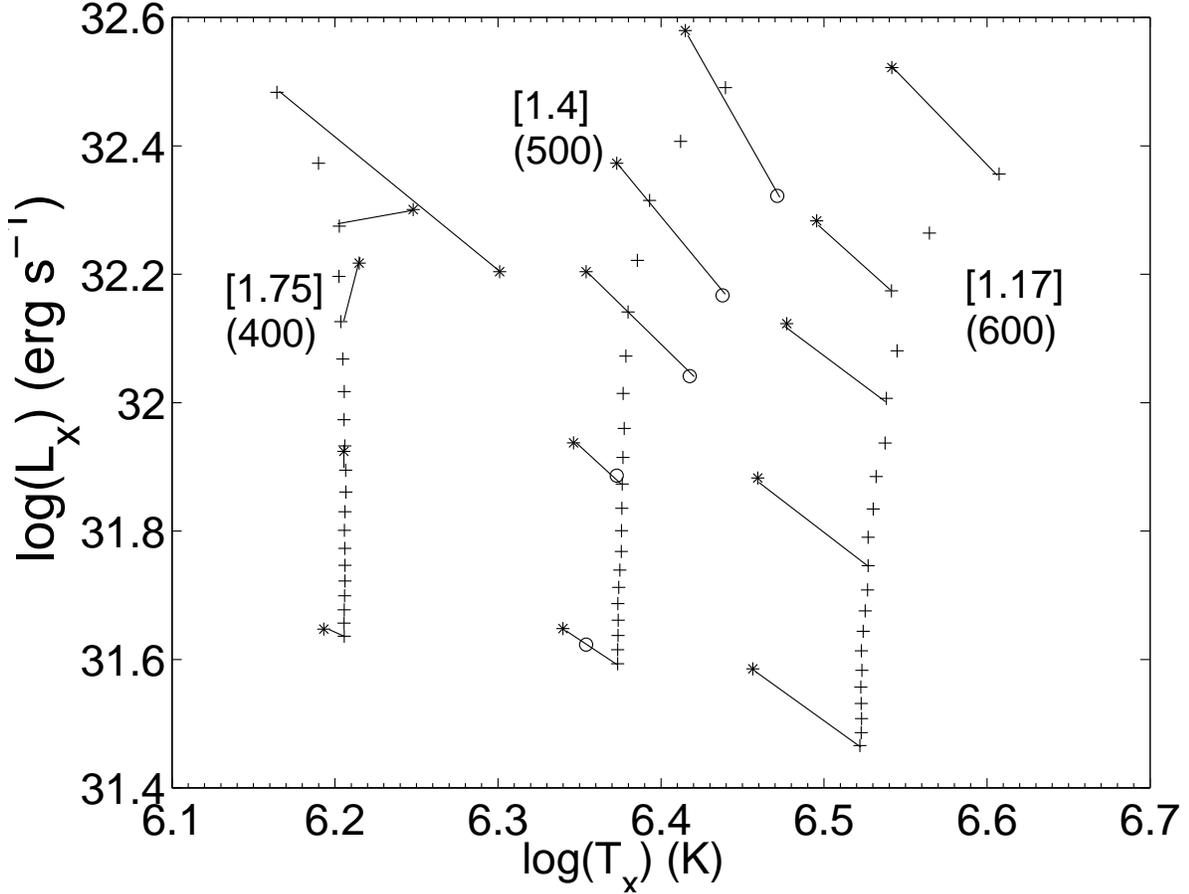}}
\caption{Comparison of PN evolution tracks in the $L_x-T_x$ plane
between the full simulations (+ signs, top to bottom  from 40 yrs
to 800 yrs, every 40 yrs) and the self similar solutions (* signs
at 40, 120, 200, 400, and 800 yrs). Radiative cooling was included
in the self similar calculations by disregarding wind segments
with $\tau _{\rm cool} < t$ for the X-ray emission. Equal-time
data points are connected by lines to facilitate the comparison.
The three cases considered differ by their $[\dot M_2/10^{-7}
M_\odot \yr^{-1}]$ and $(v_2/\km \s^{-1})$ values as indicated.
$\dot M_1 = 7 \times 10^{-6} M_\odot \yr^{-1}$ and the initial
slow wind temperature is $10 \K$ in all runs. Circles ($\circ$) in
the middle run mark the self similar results in which wind
segments with $\tau _{\rm cool} < 3t$ were disregarded. }
\label{self1}
\end{figure}

A comparison of the simulation results with the self-similar
solutions is presented in Fig.~\ref{self1} in terms of the PN
evolution in the $L_x-T_x$ plane. Three cases are shown. The fast
wind mass loss rate $\dot M_2$ in units of $10^{-7} M_\odot
\yr^{-1}$ is marked next to each track inside square brackets [ ],
while the fast wind speed in units of $\km \s^{-1}$ is in
parentheses ( ). The mass loss rate of the slow wind is $\dot M_1=
7 \times 10^{-6} M_\odot \yr^{-1}$ and the slow wind speed is
$v_1=10 \km \s^{-1}$ in all runs. Short straight lines connect the
self similar results to the numerical results at the same age $t$.
Evidently, the self similar solutions with our approximation for
radiative cooling provide very good estimates to the X-ray
properties of the nebula. In all cases presented in
Fig.~\ref{self1}, the self similar solutions reproduce $T_x$ and
$L_x$ to within $\sim 30\%$ and $\sim 45\%$, respectively, of the
simulation values. For ages of 200~yr and more, the self similar
results for $T_x$ and $L_x$ are good to within $\sim 12\%$ and
$\sim 30\%$. Recall that the simulation results at early times are
uncertain as discussed in \S2. It can be seen in the figure, that
using less strict criteria for omitting fast-cooling wind segments
(e.g., $t_{\rm cool} <3t$) improves the late-time self similar
results at the price of larger discrepancies at earlier times. For
the higher velocity simulations (two right cases in
Fig.~\ref{self1}) the radiative cooling removes cooler gas more
efficiently at early times, hence the temperature is higher than
at later times. For the slower fast wind case, $v_2=400 \km
\s^{-1}$, the post shock temperature is low and density high, such
that radiative cooling is very efficient. Because of that the
entire shocked fast wind is cooling at early times. This explains
its low temperature at early times.

In all cases, the general behavior of $L_x \propto t^{-\eta}$ with
$\eta \sim 0.9$ is obtained. This dependence on time can be
understood as follows. The velocity of the contact discontinuity
$v_c$, is constant, hence the volume $V$, of the hot bubble
increases as $V =4 \pi (R_c^3-R_2^3)/3 \sim 4 \pi (v_c t)^3/3$,
and the density decreases as $n \propto (\dot M_2 t)/V \propto
t^{-2}$. As the temperature does not change much, the emissivity
(power per unit volume) $\epsilon$ goes as $\epsilon \propto
n^2\Lambda(T) \propto n^2 \propto t^{-4}$, and $L_x =V\epsilon
\propto t^{-1}$. {{{ This simple explanation, and the close fit of
the numerical results to the self-similar solution with cooling
gas removed (Fig. \ref{self1}), enhance our confidence in our
results. We find $R_c \propto t^{0.98}$ and the density inside the
bubble to vary as $\rho \propto t^{-2}$, which gives for the mass
in the bubble $M \propto ~t^{0.94}$. As more mass cools with time,
we expect less mass to be in the bubble. }}} SS06 find in their
numerical simulations $\eta = 0.37$. We have no explanation for
this discrepancy, {{{ but we note that SS06 find the mass in their
X-ray bubble to increase as $~t^{1.5}$, rather than the simple
expectation of $\sim t$. They explain it by more efficient cooling
at early times. However, we note that this behavior continues to
late times, $t > 2000 \yr$ where the contact discontinuity is
already at $r \simeq 10^{17} \cm$. }}}

The results of the numerical simulations with constant fast wind
properties strengthen our conclusions from self similar
calculations (ASB06; see also SS06). Basically, X-ray emission
from PNs can be accounted for by shocked wind segments that were
expelled during the early PN phase, if the fast wind speed is
moderate, $v_2 \sim 400-600 \km \s^{-1}$, and the mass loss rate
is a few$\times 10^{-7} M_\odot \yr^{-1}$. More generally, the
X-ray emission from PNs can be accounted for by material ejected
at speeds of $v_2\sim 400-600 \km \s^{-1}$, whether from the
central star during the post-AGB or from a companion blowing jets
at the very late AGB phase  (Soker \& Kastner 2003). This fast
flowing gas hits the slow wind, and passes through a shock wave.
The post-shock gas is the X-ray emitter. In the next section we
study more realistic, evolving fast winds in an attempt to further
constrain the fast wind properties, which can produce observed
X-ray emission of PNs. The bipolar X-ray morphology of several
observed PNs, which indicates an important role of jets rather
than a spherical fast wind, cannot be explained by the flows
studied in this paper, and will be studied with a 2D numerical
code in a future paper.

A comment on the mass loss rate and the alternative heat
conduction model is needed here. Observations of mass loss rates
during the PN phase show very low mass loss rates, $\dot M_2 \la
10^{-8} M_\odot \yr^{-1}$, and high speeds, $v_2 \ga 1000  \km
\s^{-1}$,  (Cerruti-Sola \& Perinotto 1989; Perinotto et al.
1989). As shown by ASB06, such winds cannot account for the X-ray
emission as the luminosity will be too low and temperature too
high. This led to the idea of a heat conduction front between the
hot bubble and the cold slow wind gas (Soker 1994). The density
and temperature in the heat conduction front are intermediate,
between those of the hot bubble and the cool slow wind gas. Heat
conduction can make a significant effect and is required to
explain the X-ray emission {\it only} if after the AGB wind there
is a sharp transition to the PN-phase wind. Indeed, Zhekov \&
Perinotto (1996) in calculating the effect of heat conduction used
a fast wind mass loss rate of $\dot M_2 = 7 \times  10^{-9}
M_\odot \yr^{-1}$ for a fast wind speed of $v_2 \simeq 500 \km
\s^{-1}$. Namely, the heat conduction model is based on the
assumption that the relatively high mass loss rate ceases before
the post-AGB wind speed reaches $v_2 \sim 400 \km \s^{-1}$. This
type of wind implies a relatively long post-AGB evolution.

A different scenario would be a dense post-AGB wind, i.e., one in
which the mass loss rate stays relatively high until later times,
at least until the fast wind speed increases to $v_2 \sim 600 \km
\s^{-1}$ (Soker \& Kastner 2003; ASB06). These are the type of
winds considered here in Fig. \ref{self1}. Heat conduction, even
if it exists here, will not change much, as the density of the hot
bubble is already high enough for radiative cooling to be
important, as seen in Fig. \ref{cool1}. This can be demonstrated
as follows. At early times the pressure in the hot bubble is
larger and temperature lower that that assumed in the heat
conduction model (Soker 2004). Substituting a pressure of $n
T=3\times 10^{8} \cm^{-3} \K$ and a temperature of $T_h=2.5 \times
10^6 \K$, as appropriate near the contact discontinuity (see Fig.
\ref{cool1}), in equation (6) of Soker (1994), we find that
radiative losses become important already at an age of $t \simeq
500 \yr$. In the case of a very hot bubble ($T_h=10^7 \K$), as
assumed by Soker (1994), radiative losses dominate at much later
times, $t>10,000  \yr$.

There are several reasons to prefer the dense post-AGB wind
scenario: (1) Even weak magnetic fields, which are likely to exist
to some level in all post-AGB stars, will inhibit almost
completely heat conduction. {{{ It is still possible that mixing
occurs there (Chu et al. 1997), e.g., via turbulence (see the 2D
simulations of SS06). }}} (2) A large fraction of the X-ray bright
PNs are very young (Soker \& Kastner 2003), showing that the
transition from the AGB to the PN phase is short. This implies a
high mass loss rate during the post-AGB phase. (3) The strong
X-ray emitting PN BD +30$^\circ$3639 (PN G064.7+05.0) currently
has a wind speed of $\sim 700 \km \s^{-1}$ (Leuenhagen et al.
1996) as assumed in the dense fast wind model, and much slower
than required in the heat conduction model. However, we caution
that it is possible that the X-ray emission in this PN results
from a CFW (jets), and not from the central fast wind. To
summarize, at this stage both mechanisms in which the X-ray
emitting gas is blown by the central star should be examined, as
we cannot rule out that in some PNs the X-ray emission source is a
heat conduction front (Soker 1994; Zhekov \& Perinotto 1996;
Steffen et al. 2005). However, at this point we favor the denser
post-AGB model, which is the subject of this paper.

\section{TIME VARYING FAST WIND}
\label{evolve}

\subsection{General Considerations}
\label{general}

The conclusion of the previous section, which strengthens previous
results (Soker \& Kastner 2003, ASB06, and SS06) is that the fast
wind segments that account for the X-ray emission should have
speeds of $v_2 \simeq 400-600 \km \s^{-1}$, and the mass loss rate
should be $\dot M_2 \sim 10^{-7} M_\odot \yr^{-1}$. Thus, within a
few hundred years the mass of the relevant X-ray emitting plasma
would be a few times $\sim 10^{-5} M_\odot$. The fast wind,
however, evolves with time. In this section we assume that the
X-ray emitting gas in PNs comes from a spherically symmetric
central stellar wind, and use the X-ray properties of PNs to learn
about and constrain the evolution of the fast wind. The fast wind
evolution depends on the mass of the central star (e.g., Villaver
et al. 2002), and possibly on an interaction with a companion.
There are many uncertainties, and several different functions for
the dependence of the fast wind mass loss rate and velocity on
time have been proposed (e.g., Mellema 1994; Villaver et al. 2002;
Perinotto et al. 2004; Garcia-Segura et al. 2006). To demonstrate
the basic properties of an evolving fast wind, we consider several
types of fast wind evolution.

Since slow winds can not produce X-rays, we start our simulations
from when the fast wind speed is already $v_2 \simeq 200-300 \km
\s^{-1}$, which is after the PN stage has already started (for
example, we assume minimum temperature of $10^4 \K$ due to
ionization by the central star). When comparing to observations,
the real dynamical age of the PN, thus, is larger than the time
(age) of the simulations by several hundred years. Our goal is to
conduct a general study to find the influence of the different
parameters on the X-ray emission, rather than to fit each PN with
an appropriate mass loss evolution.

\subsection{Different Simulated Cases}
\label{scases}

The strategy of varying the wind speed is to take an initial
velocity of $v_{20} =300 \km \s^{-1}$ and to allow it to increase
gradually to currently observed speeds. The rise in velocity is
assumed to take place over a typical time $\tau_v = 1000$~yr. For
simplicity, we start by assuming that the rise in velocity is
linear.
\begin {equation}
v_2 = v_{20} \left(1+ \frac{t}{\tau_v}\right) . \label{linearv}
\end {equation}
Our different cases are summarized in Table~1.
\begin{table}

Table 1: Cases Calculated in section 4.2

\bigskip

\begin{tabular}{|l|c|c|c|c|c|c|}
\hline
Run & $\dot M_1$ & $\dot M_2$   & $v_1$ & $v_2$  & $\tau_v$  \\
&$M_\odot \yr^{-1}$ & $M_\odot \yr^{-1}$ &$\km \s^{-1}$ &
$\km \s^{-1}$ & yr\\
\hline
A & $10^{-5}$ &$3\times 10^{-7} (v_{20}/v_2) $ &$10$ &$eq. 1$ &1000 \\
\hline
B.1 &$10^{-5}$ &$3\times 10^{-7} (v_{20}/v_2)^2$ &$10$ & $eq. 1$ &1000  \\
\hline
B.2 &$10^{-4}$ &$3\times 10^{-7} (v_{20}/v_2)^2$ &$10$ & $eq. 1$ & 1000 \\
\hline
B.3 &$10^{-4}$ &$3\times 10^{-7} (v_{20}/v_2)^2$ &$10$ &$eq. 1$ &500    \\
\hline
B.4 &$10^{-4}$ &$10^{-6} (v_{20}/v_2)^2$ &$10$ & $eq. 1$&1000  \\
\hline
C &$10^{-4}$ & From P04 &10& From P04  &\_  \\
\hline
\end{tabular}

\footnotesize
\bigskip

Notes: (1) $\tau_v$ is defined in equation (1). (2) P04 refers to
Perinotto et al. (2004). \normalsize
\end{table}

We first consider two versions of wind evolution.

\begin{figure}  
\resizebox{0.95\textwidth}{!}{\includegraphics{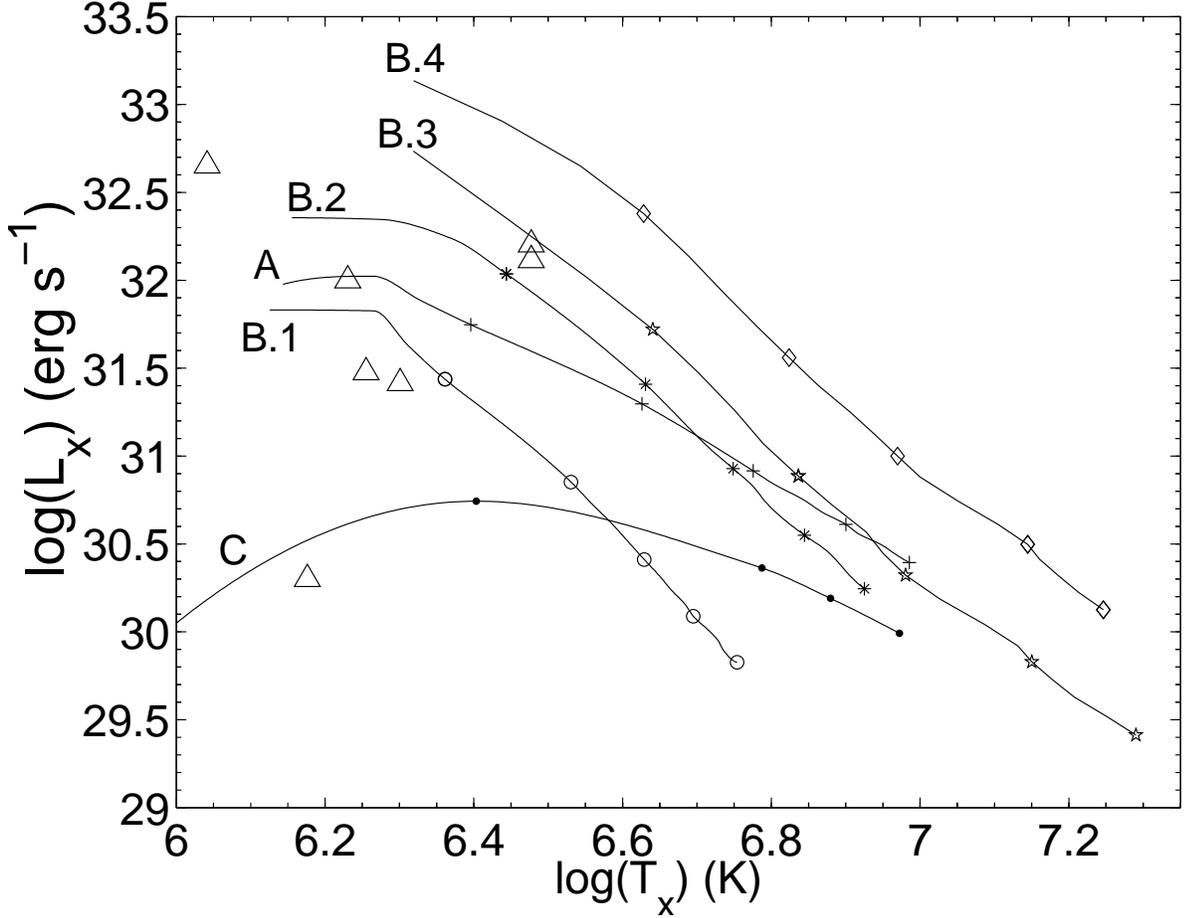}}
\caption{Evolution of X-ray emission from $t=250$~yr to
$t=5000$~yr for the evolving-wind cases described in \S
\ref{scases}. Symbols are plotted every 1000~years. Plus signs
show the constant momentum deposition case (A). Constant
luminosity cases B.1 -- B.4 are represented respectively by
circles, asterisks, stars, and diamonds. Dots show the evolution
for the wind according to Perinotto et al. (2004) for which the
$t=1000$~yr mark is beyond the scale of the graph to the left.
Triangles represent observed PNs which are listed in Table~2. All
simulations at late times predict bright, detectable, high-$T_X$
PNs that are not observed implying that the wind must evolve (i.e.
$\dot M_2(t)$ must decrease) faster than in these models. }
\label{linear}
\end{figure}

(A) Constant momentum deposition rate $\dot p_2 = \dot M_2 v_2$,
i.e. decreasing $\dot M_2 = 3 \times 10^{-7} (v_{20}/v_2) M_\odot
\yr^{-1}$. The evolution of the X-ray emission in the $L_x-T_x$
plane for this case with $\dot M_1= 10^{-5} M_\odot \yr^{-1}$ (as
in all cases here $v_1=10 \km \s^{-1}$) is shown in Fig.
\ref{linear} by the $'+'$ symbols. The results are shown from
$t=250 \yr$ to $t=5000 \yr$, with marks given every thousand
years. It can be seen that $T_x$ increases with the velocity,
while $L_x$ decreases since $\dot M_2$ decreases. The total mass
that is lost to the fast wind over $2000 \yr$ is $\Delta
M_{w2}=3.3 \times 10^{-4} M_\odot$. This is approximately the mass
available for the wind. Therefore, one could expect the wind in
this model to cease after $\sim 2000 \yr$.

(B) Constant kinetic power (luminosity) $\dot E_2=(1/2) \dot M_2
v_2^2$, which together with equation~(\ref{linearv}) implies a
decreasing mass loss rate: $\dot M_2 = \dot M_{20}
(v_{20}/v_2)^2$. The total mass lost in the fast wind in this case
is $\Delta M_{w2}= \dot M_{20} \tau_v$. We consider several cases
with constant kinetic power.

(B.1) First, we take the typical parameter values: $\dot M_2 =
3\times 10^{-7} (v_{20}/v{2})^2 M_\odot \yr^{-1}$, $\tau_v=1000
\yr$, $v_{20}=300 \km \s^{-1}$ and $\dot M_1= 10^{-5} M_\odot
\yr^{-1}$. The simulation results are represented in
Fig.~\ref{linear} by circles. The temperature $T_x$ can be seen to
decrease faster in this case than in the constant momentum
deposition case (A), because here $\dot M_2 \propto v_2^{-2}$ and
not $\dot M_2 \propto v_2^{-1}$.

(B.2) Next, we change the mass loss rate in the slow wind to $\dot
M_1= 10^{-4} M_\odot \yr^{-1}$; all other parameters and time
scales remain as in run B.1. These results are represented in Fig.
\ref{linear} by asterisks. The denser slow wind confines the hot
bubble to a smaller volume, hence leading to higher X-ray
luminosity.

(B.3) Next, we want to test a faster evolving wind with a typical
time scale of $\tau_v=500 \yr$ instead of $\tau_v=1000 \yr$
(equation (\ref{linearv})). All other parameters are as in (B.2),
e.g., the same constant wind kinetic power and same slow wind
parameters. These results are shown in Fig. \ref{linear} by star
symbols. The faster evolving wind results in a more rapid increase
in temperature at all times. At early times this leads to higher
X-ray luminosity because the post-shock fast wind is hotter,
emitting stronger in the X-ray band relative to run B.2. At later
times the luminosity is lower than in run B.2 (for equal times)
because the mass loss rate is lower in run B.3, and hence the
density inside the hot bubble is lower.

(B.4) The diamonds symbols mark a case similar to B.3, but with
denser fast wind: $\dot M_2 =  10^{-6} (v_{20}/v_2)^2 M_\odot
\yr^{-1}$ instead of $\dot M_2 = 3 \times 10^{-7} (v_{20}/v_2)^2
M_\odot \yr^{-1}$. As expected, the denser bubble has a higher
luminosity.

(C) Finally, we wish to test the wind evolution proposed by
Perinotto et al. (2004; also Sch\"onberner et al. 2005a,b). Taking
$v_2(t)$ and $\dot M_2(t)$ from Fig. 1 (Sch\"onberner et al.
2005a,b) we obtain the results represented in Fig.~\ref{linear} by
the dots. Here $v_2(t)$ is very low until $t=250$~yr, so we do not
see much X-ray until after 1000~yr, hence the 1000~yr mark is not
on this plot.

\subsection{Comparison with Observations and Discussion}
\label{fastw}

The various simulations are compared in Fig.~\ref{linear} to seven
observed PNs whose details are given in Table~2. It can be seen
that a linear wind-speed evolution could generally explain the
bright PNs during the early phase of
 $t \la 1000 \yr$.
For the brightest PNs, the slow wind needs to be dense, $\dot M_1
\sim 10^{-4} M_\odot \yr^{-1}$ as assumed in runs B.2 -- B.4.
Also, the high temperatures of $\log T_X > 6.5$ produced in all of
the simulations at late times occur when the luminosity is
decreasing, but still at a detectable level. However, to date no
PN has been detected at these high X-ray temperatures. If there
are PNs with higher temperatures, they must be as faint as the
faintest PNs in Table~2 or fainter. This implies that the fast
winds must evolve faster than assumed in our simulations, so that
by the time the temperature rises beyond $\log T_X = 6.5$, $\dot
M_2 $ is too low to produce significant X-rays.

SS06 found a similar result, although they examined only one case
of fast wind evolution. They conclude that the shocked fast wind
cannot account for X-ray properties of BD~+30$^\circ$3639 and NGC
40. We disagree with SS06 on that matter, as a more rapidly
evolving fast wind can in fact account for the X-ray properties.
In \S \ref{late} we explore this possibility.


\begin{table}

Table 2: X-ray properties of Planetary Nebulae

\bigskip
\begin{tabular}{|l|c|c|c|c|}
\hline
\#&PN & $L_x$ & $T_x$ & Dynamical Age \\
 &  & $ 10^{32}$ erg s$^{-1}$ & $ 10^{6}$ K & yr\\
\hline
1& NGC 7027(PN G084.9-03.4) &1.3 &3 &600 \\
\hline
2& BD +30 3639(PN G064.7+05.0) &1.6 &3&700 \\
\hline
3&NGC 7026(PN G096.4+29.9) & 4.5 &1.1 &$<1000$  \\
\hline
4&NGC 6543(PN G096.4+29.9) &1.0&1.7 &1000  \\
\hline
5 &NGC 7009(PN G037.7-34.5) &0.3 &1.8 &1700\\
\hline
6 &NGC 2392 (PN G197.8+17.3) &0.26 &2 & 1800\\
\hline
7 &NGC 40 (PN G120.0+09.8) &0.02 &1.5 & 5000\\
\hline
\end{tabular}

\footnotesize
\bigskip
The parameters of the first three PNs and NGC 7009 are summarized
by Soker \& Kastner (2003). The data for NGC 2392 are from
Guerrero at al.\ (2005), for NGC 40 from Montez et al.\ (2005),
and for NGC 7026 from Gruendl et al. (2006). \normalsize
\end{table}

The fast wind evolution as used by Perinotto et al. (2004) results
in low luminosities inconsistent with six out of the seven
observed PNs. The lowest-$L_X$ PN, NGC~40, might be explained by
this wind. However, the low temperature observed for this target
$T_x \simeq 1.5 \times 10^6 \K$ occurs in the Perinotto et al.
(2004) wind at $\sim$~1500~yr, while NGC~40 is believed to be
5000~yrs old.

According to Perinotto et al. (2004), the fast wind speed rises
from $200 \km \s^{-1}$ to $700 \km \s^{-1}$ in $1300 \yr$. This
appears to be too slow. For example, the typical PN
BD$+30^\circ3639$ has a dynamical age of only $\sim 800 \yr$ (Li
et al. 2002), but its fast wind speed is already $\sim 700 \km
\s^{-1}$ (Leuenhagen et al. 1996), and its mass loss rate is much
higher than that used by Perinotto et al. (2004) This suggests
that the post-AGB evolution of strong X-ray emitting PNs is faster
than what single star models (Perinotto et al. 2004) predict, but
it is also possible that the X-ray emission in BD$+30^\circ3639$
and other similar PNs result from a CFW (jets), and not from the
central fast wind of a single star.

There are more reasons to believe rapid pre-PN (post-AGB)
evolution. Recent studies suggest that the majority of observed
PNs are descendant of binary systems (Moe \& De Marco 2006; Soker
\& Subag 2005; Soker 2006). The stellar companion not only is
expected to shape the AGB wind such that the descendant PN be
axi-symmetrical rather than spherical, but the companion is also
likely to disturb the AGB progenitor such that the mass loss rate
will be much higher (Soker \& Subag 2005; Soker 2006). The
disturbed AGB envelope will affect the post-AGB phase, so that the
post-AGB phase will be shorter than that expected from single star
evolution (Soker 2006). This consideration is another
justification for invoking a rapidly evolving fast wind (\S
\ref{late}) as eventually the mass that is left on the central
star is very low, and the mass loss rate must decrease fast,
reducing the kinetic power in the wind.

\subsection{Modified Late-Phase Evolution}
\label{late}

Given the inappropriateness of the previous models at late times,
in this section we consider a different late-stage evolution
scenario in which at some point the mass loss rate starts
decreasing more rapidly than $\dot M_2 \sim v_2^{-2} \sim t^{-2}$.
We take the fast wind velocity to continue increasing linearly
according to equation \ref{linearv}, but at time $\tau _s$ the
wind's kinetic power $\dot E_2$ is no longer constant, and it
starts to decrease linearly:
\begin{equation}
\frac{\dot E_2}{\dot E_{20}} =
\begin{cases}
 1  & \qquad  0 \le t < \tau_s
 \\
 1- (t-\tau_s)\tau_d^{-1}   & \qquad  \tau_s  \le t ,
\end{cases}
\label{power1}
\end{equation}
where $\tau _d$ is the typical decay time. We also set a minimum
mass loss rate $\dot M_{\rm 2min}$.

\begin{figure}  
\resizebox{0.95\textwidth}{!}{\includegraphics{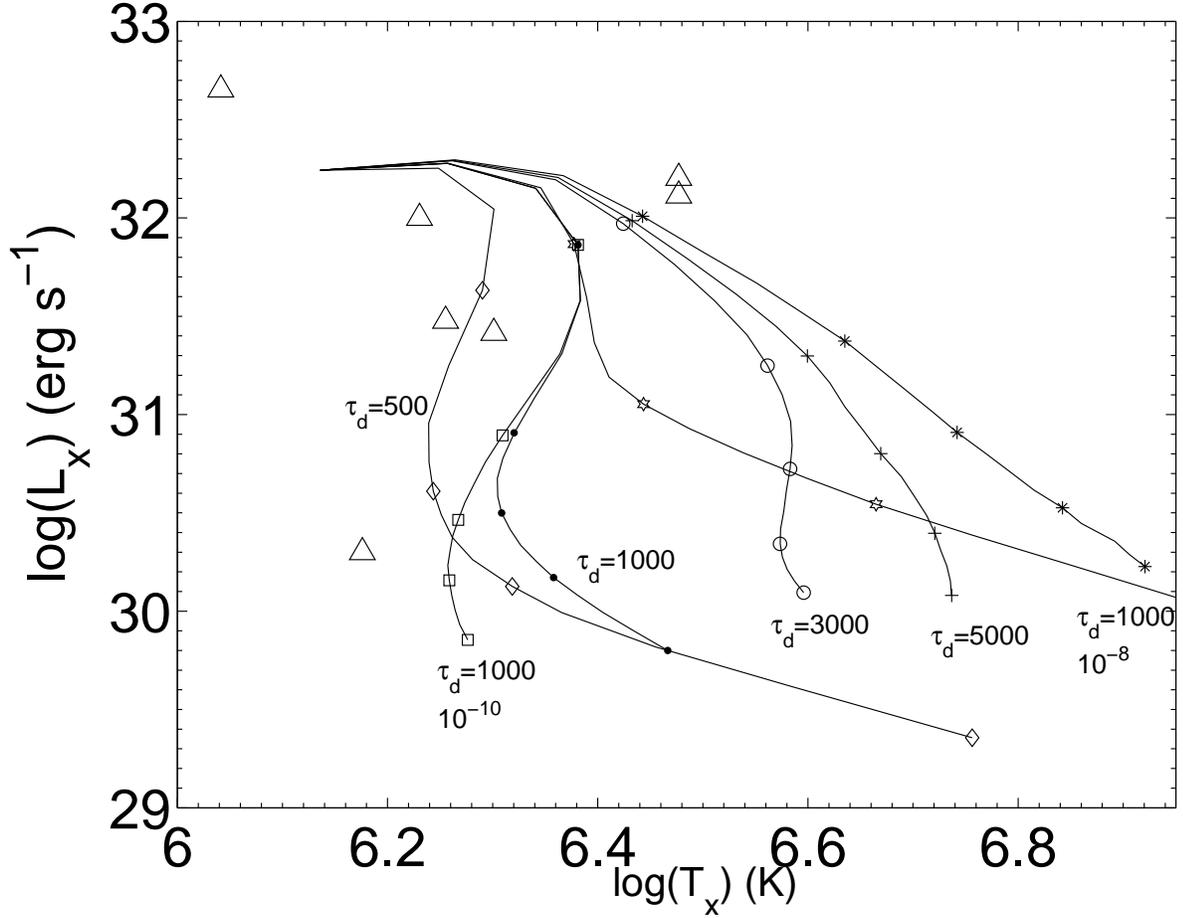}}
\caption{Evolution of the X-ray emission for a rapidly decaying
fast wind as in equation~\ref{power1}. The value of $\tau_d$ is
marked near each run. In all runs $\dot M_1= 10^{-4} M_\odot
\yr^{-1}$, $\tau_v=1000 \yr$, and $\tau_s=500 \yr$. $\dot M_{\rm
2min}= 10^{-9}M_\odot \yr^{-1}$ in all cases except the one marked
$10^{-10}$ and the one marked $10^{-8}$. Symbols mark 1000~yr
intervals, up to 4000~yr in the run marked with diamonds, and up
to 5000~yr in all other runs. All plots start at 250~yr. }
\label{late1}
\end{figure}

In the first run, we take case B.2, namely $\dot M_1= 10^{-4}
M_\odot \yr^{-1}$ and $\tau_v=1000 \yr$ at early times and for
late times using equation~\ref{power1}, we take $\tau_s=500 \yr$,
and $\tau_d=5000 \yr$, and $\dot M_{\rm 2min}= 10^{-9}M_\odot
\yr^{-1}$. In the following runs, we vary $\tau_d$. The results
are shown in Fig. \ref{late1}, where we also show one run with
$\dot M_{\rm 2min}= 10^{-10}M_\odot \yr^{-1}$ and one with $\dot
M_{\rm 2min}= 10^{8} M_\odot \yr^{-1}$, both of them for
$\tau_d=1000 \yr$.

The results of Fig.~6 can be interpreted as follows. As the supply
of gas decreases, the hot bubble density decreases because of
expansion, and thus $L_X$ decreases. The energy budget of the hot
bubble then is dominated by adiabatic cooling. This explains why
the temperatures are lower than in the previous section, although
the evolution of the wind velocity $v_2(t)$ remained unchanged.
This type of flow might result from very rapid post-AGB evolution,
or if the confinement of the hot bubble by the cold shell is not
perfect; For example, the two jets in NGC 40 (Meaburn et al. 1996)
through which hot gas can escape, and thus reduce the X-ray
temperature and luminosity.

\begin{figure}  
\resizebox{0.95\textwidth}{!}{\includegraphics{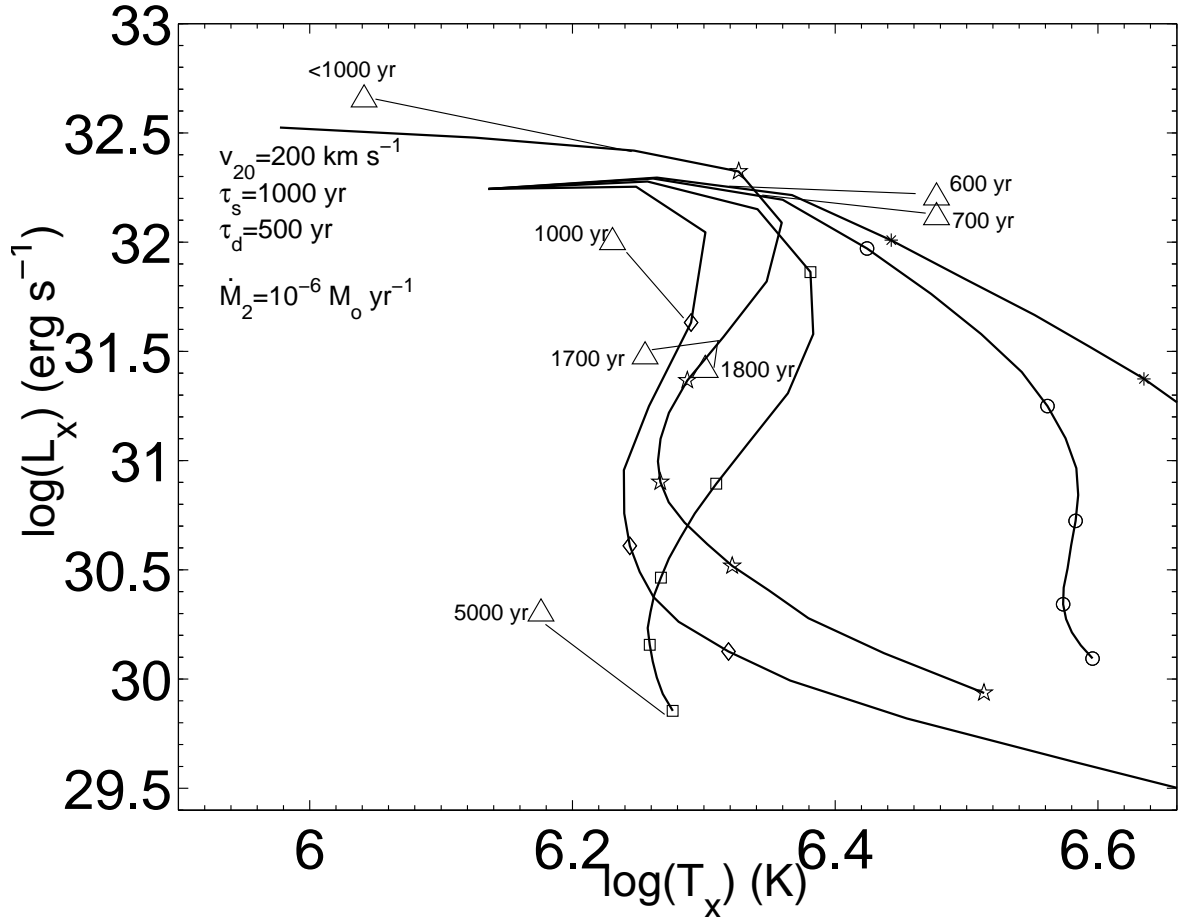}}
\caption{As Fig. \ref{late1}, with three runs omitted. The
dynamical age of PNs (in years)are marked, and each PN is
connected to a wind age equals to its dynamical age. The stars
marked a new run with parameters as written on the upper left
corner. } \label{late2}
\end{figure}

In Fig. \ref{late2} we connect the different PNs to the nearest
plot having its age equal to the dynamical age of the PN. The
dynamical age of the PNs are marked on the figure. We removed
three runs and added one run where we started the fast wind with
$v_{20}=200 \km \s^{-1}$, and took $\tau_s=1000 \yr$, $\tau_d=500
\yr$, $\dot M_{20}=10^{-6} M_\odot \yr^{-1}$, and $\dot
M_1=10^{-4} M_\odot \yr^{-1}$ (star symbols). This figure shows
that in addition to the luminosity and the temperature of the
X-ray emitting gas, we can account for the age of PNs as well.

Since we start the simulations at the PN stage, the PN time $t$ in
the simulation is somewhat smaller than the actual age of the PN,
by $\sim few \times 100 \yr$. Therefore, in Fig.~\ref{late2}, we
connected the observed PNs to simulated $t$ values that are
somewhat smaller than the observed age of the PN. For example, the
upper left PN whose estimated age is 1000~yr is connected by a
line to a $t= 750 \yr$ simulation data point. We mention again
that our goal is to find the general constraints on the properties
of the fast wind, if it is the source of the X-ray emission,
rather than to fit individual PNs. Changing a little the
parameters used here, we can fit each PN individually. In
addition, the composition of the fast wind also affects the X-ray
properties (SS06); we use constant (solar) composition for the
cooling function.

\begin{figure}  
\resizebox{0.95\textwidth}{!}{\includegraphics{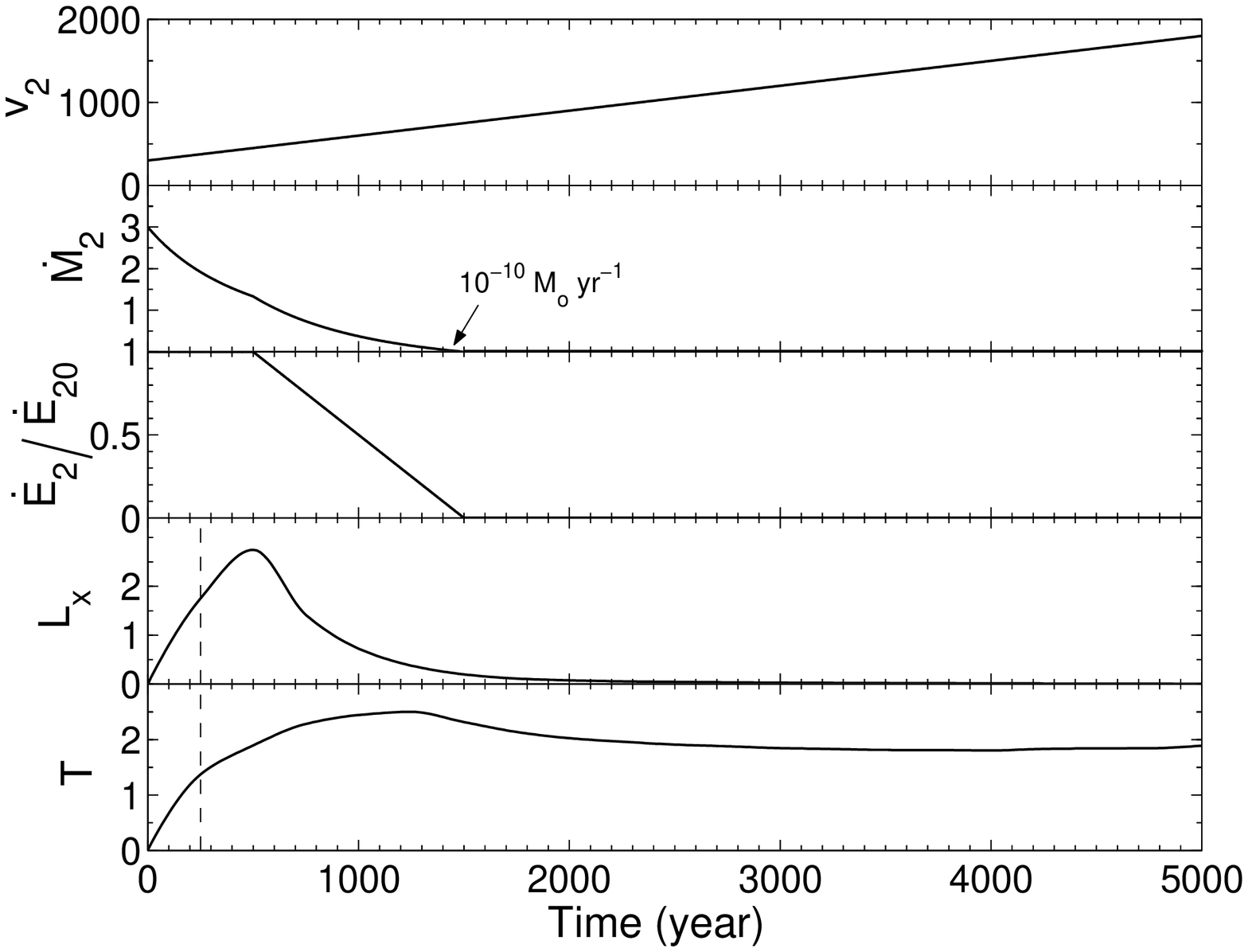}}
\caption{From top to bottom, the evolution of the fast wind speed
in $\km \s^{-1}$, the mass loss rate in $10^{-7} M_\odot
\yr^{-1}$, the logarithm of the ratio of kinetic power to its
initial value, the X-ray luminosity in $10^{-32} \erg \s^{-1}$ and
temperature in $10^6 \K$, for the run marked by squares in Figs.
\ref{late1} and \ref{late2}). The first point of each run shown on
Figs. \ref{late1} and \ref{late2} is at an age of 250 years, as
marked here by the dashed vertical line. Note that because of
radiative cooling of cooler gas, $T_x$ increases at late times. }
\label{windd1}
\end{figure}

We finish this section by showing the evolution with time of the
most important physical quantities. In Fig.~\ref{windd1} we show
the evolution with time of the quantities in the run with  $\dot
M_1= 10^{-4} M_\odot \yr^{-1}$, $\tau_v=1000 \yr$, $\tau_s=500
\yr$, $\tau_d=1000 \yr$, and $\dot M_{\rm 2min}= 10^{-10}M_\odot
\yr^{-1}$. The total mass lost in the fast wind (from $t=0$ in our
calculation, when $v_2=300 \km \s^{-1}$) at 500, 1000, and 1500
years is $\Delta M_2 = 10^{-4} M_\odot$,  1.387$\times 10^{-4}
M_\odot$, and $ 1.4675\times 10^{-4} M_\odot$, respectively. {{{
Initially, the reverse (inner) shock moves outward, as does the
contact discontinuity. At $t=1250 \yr$ it starts to move inward
because the ram pressure of the fast wind decreases by a large
factor. At $t=1750 \yr$ it starts to move outward again. }}}

\section{SUMMARY }

Our main finding is that if the source of the extended X-ray
emitting gas in PNs is the fast wind blown by the central star,
then the fast wind must be rapidly evolving. Otherwise the
luminosity would be too low at early times, and the fast wind
would produce hot gas at later times that is not observed. Our
simulations require that the mass loss rate of the fast wind be
relatively high, $\dot M_2 = 1-3 \times 10^{-7} M_\odot \yr^{-1}$,
at early times when the wind speed is $\sim 300-700 \km \s^{-1}$,
and then it needs to rapidly decline to (Fig. \ref{windd1}). This
weakening of the wind occurs approximately $500-1500 \yr$ after
the fast wind has started to blow, or $\sim 1000-3000 \yr$ after
the star has left the AGB. Our results are contradictory to those
of SS06, who examined only one case of fast wind evolution, and
could not find a match between their model and the X-ray
properties of BD~+30$^\circ$3639 and NGC 40.

A rapid fast-wind evolution is compatible with a recent claim for
rapid post-AGB evolution caused by binary interaction (Soker
2006). The high mass loss rate at the early PN phase is a
continuation of a high mass loss rate during the final AGB phase.
Such a high mass loss rate may lead to dust formation and strong
IR emission. Montez (2006; R. Montez, private communication)
suggested a correlation between IR emission and X-ray emission.
Not all PNs will have such rapid fast-wind evolution, hence not
all PNs will have detectable X-ray emission. Other PNs, those with
slowly evolving fast winds are not expected to produce detectable
X-ray emission as adiabatic cooling overcomes shock heating. {{{
Note that a slowly evolving fast wind requires a low mass loss
rate. The reason is that during the post-AGB phase the mass in the
envelope is very low, and the star cannot lose mass at a high rate
for a long time. The low mass lose rate implies a weak X-ray
emission, as the case C in Figure \ref{linear}. }}}

We should stress here that we are not claiming that {\it all} PNs
owe their X-ray emission to the central star fast wind.
Definitely, in some cases jets are the source of the X-ray
emitting gas. If observations indicate that the fast wind
evolution in PNs is much slower than required by our models,
(e.g., Perinotto 2004), we will have to agree with SS06 and
conclude that a central star wind can not explain the observed
X-rays. Jets blown by a companion may be an alternative
explanation for the X-ray emission. In this case we also expect
that only a fraction of all PNs will have detectable X-ray
emission. Heat conduction can also play a role, although our
models show that heat conduction is not required if the fast wind
evolves fast enough.

As both rapid fast-wind evolution and jets require the presence of
a close companion (in the case of jets a stellar companion, in the
case of rapid evolution a brown dwarf or even a massive planet
might do the job as well), we propose that the extended X-ray
emission in PNs is tightly connected to the presence of a close
binary companion.

\acknowledgments We thank Raghvendra Sahai, Matthias Stute, Martin
Guerrero, Joel Kastner, {{{ and an anonymous referee for useful
comments. }}} This research was supported in part by the Israel
Science Foundation, grant 28/03, and by the Asher Fund for Space
Research at the Technion.

\end{document}